\documentclass[12pt]{article}
\usepackage{amsfonts,amsmath}
\usepackage{amssymb}
\usepackage{ytableau}

\newcommand{\dr}{{{\rm d}}}

\makeatletter \@addtoreset{equation}{section} \makeatother

{\vspace{3mm} }

\def\al{\alpha}

\def\*{\star}

\def\E2{\mathbf{E}}

\def\y{\mathbf{y}}
\def\v{\mathbf{v}}
\def\u{\mathbf{u}}

\def\pp{\mathtt{p}}
\def\qq{\mathtt{q}}

\newcommand{\be}{\begin{equation}}
\newcommand{\ee}{\end{equation}}
\newcommand{\bee}{\begin{eqnarray}}
\newcommand{\beee}{\begin{array}}
\newcommand{\eee}{\end{eqnarray}}
\newcommand{\eeee}{\end{array}}
\newcommand{\gb}{\beta}
\newcommand{\gga}{\gamma}

\newcommand{\gd}{\delta}

\newcommand{\gep}{\epsilon}

\newcommand{\go}{\omega}

\newcommand{\nn}{\nonumber}

\newcommand{\p}{\partial}

\newcommand{\ff}{\frac}
\newcommand{\rmx}{{\mathrm{x}}}
\newcommand{\rmz}{{\mathrm{z}}}

\begin{document}

\begin{flushright}
FIAN/TD/17-2023\\
\end{flushright}

\vspace{0.5cm}
\begin{center}
{\large\bf Toward higher-spin symmetry breaking in the bulk}

\vspace{1 cm}

\textbf{V.E.~Didenko and A.V.~Korybut}\\

\vspace{1 cm}

\textbf{}\textbf{}\\
 \vspace{0.5cm}
 \textit{I.E. Tamm Department of Theoretical Physics,
Lebedev Physical Institute,}\\
 \textit{ Leninsky prospect 53, 119991, Moscow, Russia }\\

\par\end{center}

\begin{center}
\vspace{0.6cm}
e-mails: didenko@lpi.ru, akoribut@gmail.com \\
\par\end{center}

\vspace{0.4cm}

\begin{abstract}
\noindent We present a new vacuum of the bosonic higher-spin gauge
theory in $d+1$ dimensions, which has leftover symmetry of the
Poincar\'{e} algebra in $d$ dimensions. Its structure is very
simple: the space-time geometry is that of $AdS$, while the only
nonzero field is a scalar. The scalar extends along the
Poincar\'{e} radial coordinate $\rmz$ and is shown to be linearly
exact for an arbitrary mixture of its two $\Delta=2$ and
$\Delta=d-2$ conformal branches. The obtained vacuum breaks the
global higher-spin symmetry leading to a broken phase that lives
in the Minkowski space-time.
\end{abstract}

\section{Introduction}

Higher-spin (HS) gauge theories \cite{Bekaert:2022poo} are often
thought of as underlying string theory in its allegedly unbroken
symmetry phase \cite{Gross:1988ue}; see also
\cite{Sundborg:2000wp, Bianchi:2003wx} for further related ideas.
A proposal in \cite{Metsaev:1999ui} suggested that superstrings
propagate at the boundary of 11-dimensional $AdS$
space\footnote{Notice, however, that there are no simple
supergroups in $AdS$ in $d\geq 8$, \cite{Nahm:1977tg}} as a result
of spontaneous HS symmetry breaking. The specific mechanism
relating the two theories is not practically available for a
number of reasons. For one, while an HS candidate rich enough to
embrace arguably all stringy states was recently proposed
\cite{Vasiliev:2018zer}, the analysis of this theory is still
conceptually and technically challenging even at the linearized
level (see \cite{Degtev:2019enl} for a work in this direction).
Second, HS theories formulated naturally in $AdS$ space
\cite{Fradkin:1986ka} have recently faced the locality problem
\cite{Bekaert:2015tva, Sleight:2017pcz} which still has not been
fully resolved beyond cubic order (see \cite{Roiban:2017iqg,
Ponomarev:2017qab, Neiman:2023orj, Gelfond:2023fwe} for various
approaches at quartic order).

Some accessible nonlinear HS models suitable for the
symmetry-breaking studies are available in the form of Vasiliev's
generating equations \cite{Vasiliev:1992av, Vasiliev:2003ev}. They
describe interactions of totally symmetric gauge fields at the
level of the classical equations of motion. Although the spectra
of these models are much poorer than those that string theory
suggests, the details of symmetry breaking that lead to massive
states are not known even for these simpler models. It is curious
to note in this regard how under an {\it ad hoc} assumption on the
$AdS$ symmetry breaking one arrives precisely at the stringy Regge
leading trajectory \cite{Metsaev:1999kb}.

In this paper, we attempt to make a step in a similar direction by
addressing the following simple question. Is there an HS vacuum of
the $(d+1)$-dimensional theory that has Poincar\'{e} algebra as
the global space-time symmetry in $d$ dimensions? Such a vacuum,
if it exists, provides arguably a systematic way to analyze the HS
broken phase at least in some toy model examples. We answer this
question in the affirmative by manifestly constructing the
corresponding solution of the bosonic HS equations.

With such a vacuum, one can consider perturbation theory about it.
Fluctuating fields naturally acquire dependence on the $AdS$
boundary coordinates $\vec\rmx$, as well as on the radial bulk
direction $\rmz$. This way one arrives at $d$-dimensional
(generally massive) excitations propagating in Minkowski
space-time at a fixed slice $\rmz$. Among these of great interest
are those that either depend on $\rmz$ trivially (for example, in
a scaling fashion) or result in a reorganization of HS modules
that effectively makes dynamics $d$-dimensional, along the lines
of \cite{Vasiliev:2012vf, Diaz:2024iuz}. The latter correspond to
the broken phase of $d+1$ HS theory in $d$ dimensions.

In approaching this problem, we use the standard unfolded
formalism of \cite{Vasiliev:1988sa} (see also
\cite{Misuna:2022cma} for its quantum extension) that allows one
to cast a highly nontrivial many-derivative HS interaction into
the form of first-order conditions at the cost of introducing of
infinitely many auxiliary fields. The schematic form of these
equations is
\begin{align}
&\dr_x\go+\go*\go=\Upsilon(\go,\go, C)+\Upsilon(\go,\go, C,
C)+\dots\,,\label{eqw}\\
&\dr_x C+\go*C-C*\pi(\go)=\Upsilon(\go, C, C)+\Upsilon(\go, C, C,
C)+\dots\,.\label{eqC}
\end{align}
Further details on the above system will be provided shortly. For
now, we would like to focus on its general structure. The fields
$\go=\go(Y|x)$ and $C=C(Y|x)$ are the generating functions of HS
gauge fields and their field strengths, respectively. The
generating variables, collectively called $Y$, encode spinning
components and the necessary auxiliary fields organized in
accordance with the HS algebra generated by the star product $*$.
In particular, the field spectrum contains a scalar associated
with the lowest component of $C$,
\be
\phi(x):=C(Y|x)\Big|_{Y=0}\,.
\ee
Infinite series of $\Upsilon$'s govern nonlinear gauge-invariant
field interactions. Their explicit form is at the core of the HS
problem. These can be extracted from the Vasiliev equations modulo
a field redefinition \cite{Vasiliev:2003ev}. Although systematic,
the procedure is substantially involved in practice and draws one
into the order-by-order factorization of the trace ideal, a
routine that sets the equations on shell. Given the highly
nonlinear nature of the Vasiliev theory, it is not too surprising
that there are only a handful of exact solutions available in the
literature: (\cite{Prokushkin:1998bq, Iazeolla:2015tca} in three
and \cite{Sezgin:2005pv, Sezgin:2005hf, Iazeolla:2007wt,
Didenko:2009td, Iazeolla:2011cb, Iazeolla:2017vng, Aros:2017ror}
in four dimensions; see also review \cite{Iazeolla:2017dxc}),
while there are none in arbitrary $d$ except for the trivial one
corresponding to an empty $AdS$ space,
\be\label{stnd}
\go_0=W_{AdS}\,,\qquad C_0=0\,.
\ee
Even though Vasiliev's system describes full nonlinear dynamics,
it is not yet clear which choice of field variables leads to the
vertices $\Upsilon$ within a proper class of (non)local
interactions. This problem is currently under active
investigation; see, e.g., \cite{Gelfond:2019tac, Didenko:2022eso,
Vasiliev:2023yzx}.

We do not pursue the analysis of the original equations from
\cite{Vasiliev:2003ev} in our work. Instead, we use the recently
proposed Vasiliev-like system \cite{Didenko:2022qga} specialized
to HS interactions of symmetric fields in any dimensions
\cite{Didenko:2023vna}. The advantage of the latter approach is
its manifest all-order (off-shell) locality, which clears the way
for an unexpectedly simple nontrivial vacuum of the theory.

Let us briefly comment on the difference between the original
generating equations of \cite{Vasiliev:2003ev} and those of
\cite{Didenko:2023vna}. Both systems describe unconstrained, i.e.,
off-shell nonlinear bosonic HS fields in arbitrary dimensions.
Both operate with the same set of fields governed by the off-shell
HS algebra and as such result in the same unfolded equations
\eqref{eqw}, \eqref{eqC}. The key difference is the type of the
large $(z, Y)$-algebra featuring in the generating systems, which
is responsible for the explicit form of the vertices that show up
on the right-hand sides of \eqref{eqw}, \eqref{eqC}. In the case
of Vasiliev, the large algebra contains noncommuting $z$'s, while
in our case these $z$'s commute, which is not feasible for the
generating equations of \cite{Vasiliev:2003ev} due to unavoidable
star-product divergences. Nevertheless, the commuting $z$-algebra
has already effectively come out in the analysis of
\cite{Didenko:2019xzz}, where the requirement of locality for the
Vasiliev vertices was imposed. To make it work within
\cite{Didenko:2023vna} required revising the basic elements of the
original Vasiliev equations. In \cite{Didenko:2022qga} it was
shown that the modification of the Vasiliev construction that
arises in the $z$-commuting limit is indeed possible for the $4d$
HS system. This result was then extended to any $d$ in
\cite{Didenko:2023vna}. At the level of vertices in \eqref{eqw},
\eqref{eqC} we believe the two systems from \cite{Vasiliev:2003ev}
and \cite{Didenko:2023vna} should reproduce identical results.
This can be checked at the first few interaction orders, but not
yet at higher orders, because the locality of the original
equations of \cite{Vasiliev:2003ev} is not yet settled at higher
orders.

It should be stressed once again that we are dealing with the
off-shell system here.  The HS on-shell dynamics can be obtained
using the factorization procedure, the details of which are
currently under development . Taking the quotient comes along with
the very definition of HS physical fields.

Our main finding is very simple. The HS theory of symmetric fields
in $d+1$ dimensions parametrized by the Poincar\'{e} coordinates
$x^{\mu}=(\vec\rmx, \rmz)$ has the following exact solution of
\eqref{eqw} and \eqref{eqC}:
\be\label{vac}
\go_0=W_{AdS}\,,\qquad \phi(\vec\rmx,
\rmz)=\nu_1\,\rmz^2+\nu_2\,\rmz^{d-2}\,,
\ee
where $W_{AdS}$ is the appropriately chosen $AdS_{d+1}$ connection
and $\nu_{1,2}$ are arbitrary parameters. Unlike the standard HS
vacuum \eqref{stnd}, vacuum \eqref{vac} introduces a nonzero
scalar profile, which is independent of the boundary coordinates
$\vec\rmx$. It depends on the radial $\rmz$ satisfying the
Klein-Gordon equation
\be
\Box_{AdS_{d+1}}\phi=m^2\phi\,,\qquad m^2=2(2-d)\,,
\ee
where the mass-like term is given in terms of the negative
cosmological constant. We thus show that the linearized
approximation turns out to be all-order exact, leading to
\be\label{HO}
\Upsilon(\go_0, \go_0, C_{\phi},\dots, C_{\phi})=\Upsilon(\go_0,
C_{\phi},\dots, C_{\phi})=0\,.
\ee
The proposed vacuum breaks the global HS symmetry down to a
subalgebra that has the Poincar\'{e} as the space-time symmetry in
$d$ dimensions. That the scalar does not contribute to HS sectors
is a pure kinematics. Being $\vec\rmx$ independent, it is too
symmetric, and thus offers no spin structure whatsoever. Less
clear is the absence of its nonlinear self-interaction as the
nature of the observed higher-order cancellations remains obscure
to us. On the other hand, the solution obtained is remarkably
simple and naturally suggests proceeding with the linearized
analysis about it. The corresponding free theory arguably lives on
the Minkowski background in $d$ dimensions. We hope to report on
progress in this direction elsewhere.

The remainder of this paper is structured as follows. In Section
\ref{secEq} we provide a brief review of Vasiliev's HS algebra in
$d+1$ dimensions and present the generating equations of
\cite{Didenko:2023vna}. A suitable ansatz, as well as the solution
of equations of motion is given in Section \ref{secSol}, where we
also provide the on-shell condition used, elaborate on its global
symmetries, and lay out a few basic properties of the obtained
vacuum. Our conclusions are given in Section \ref{secCon}.

\section{HS generating equations}\label{secEq}

Equations \eqref{eqw}-\eqref{eqC} contain the 1-form $\go(Y|x)$
and 0-form $C(Y|x)$ valued in an HS algebra. Following
\cite{Vasiliev:2003ev}, the bosonic HS algebra in $d+1$ dimensions
can be generated using a set of oscillators
\be
Y=(y_{\al}, \y^a_{\gb})\,,\qquad a=0\dots d\,,\qquad \al,
\gb=1,2\,,
\ee
where $a$ is the $o(d,1)$ Lorentz index, while $\al$ and $\gb$ are
attributed to an $sp(2)$, which is designed to generate two-row
Young diagrams. Indeed, as shown in \cite{Vasiliev:2003ev},
whenever the greek indices are contracted with the $sp(2)$
canonical form $\gep_{\al\gb}=-\gep_{\gb\al}$ forming an $sp(2)$
singlet, the coefficients of a polynomial $f(\y, y)$ are nothing
but a bunch of (Lorentz traceful) two-row diagrams. For example,
$\go(Y|x)$ generates the following set of HS fields of arbitrary
spin $s\geq 1$
\be\label{w}
\go^{a(s-1),\, b(n)}=\dr x^{\mu}\go^{a(s-1),\,
b(n)}_{\mu}=\ytableausetup{mathmode, boxsize=1.25em}
\begin{ytableau}
{} & {\bullet} & {\bullet}
& {\bullet} & {} \\
{} &  {\bullet} & {\bullet} & {}
\end{ytableau}\,\,,\qquad 0\leq n\leq
s-1\,,
\ee
where by $a(m)$ we traditionally denote the symmetrization over
$m$ indices. The Moyal star defines a product in the associative
HS algebra
\be\label{moyal}
(f*g)(y, \y)= \int f\left(y+u, \y+\u\right) g\left(y+v,
\y+\v\right)
e^{iu_{\al}v^{\al}+i\u^{a}_{\al}\v^{b}_{\gb}\eta_{ab}\gep^{\al\gb}}\,,
\ee
where the functions $f$ and $g$ are assumed to be $sp(2)$
singlets, while $\eta_{ab}$ is the $o(d,1)$ metric.

The generating 0-form $C(Y|x)$ from \eqref{eqw}-\eqref{eqC}
manifests the so-called twisted-adjoint module of the HS algebra,
where the twist in \eqref{eqC} is defined as the following
reflection:
\be\label{pi}
\pi f(y, \y)=f(-y, \y)\,.
\ee
The traceful two-row Young diagrams belong to what can be referred
to as the {\it off-shell} HS algebra. It contains a greater set of
fields than is required for the on-shell dynamics.
Correspondingly, the system \eqref{eqw}-\eqref{eqC} does not
describe dynamical evolution; rather, it offers a set of the
generalized Bianchi consistency constraints and conditions that
express any particular auxiliary field in terms of space-time
derivatives of other fields. Such type of unfolded equations is
usually called off shell (see, e.g., \cite{Vasiliev:2005zu,
Grigoriev:2006tt}). The on-shell spectrum contains fewer fields.
Namely, those associated with the Lorentz traces of the two-row
Young diagrams have to be consistently dismissed. A proper way of
doing this is via factorization of the trace ideal. The reader may
find more on this matter in \cite{Bekaert:2004qos}. Let us also
add to this: star products in \eqref{eqw}-\eqref{eqC} do not
respect a chosen on-shell field representative condition in
general. Thus, Eqs. \eqref{eqw}-\eqref{eqC} should be treated
modulo terms from the corresponding ideal.

\paragraph{Generating equations}
Vertices on the right-hand sides of \eqref{eqw}-\eqref{eqC} can be
generated order by order using equations from
\cite{Didenko:2023vna}, which are based on the Vasiliev idea that
$\go(Y|x)$ can be embedded into a bigger space with the extra two
coordinates $z_{\al}=(z_1, z_2)$
\be
W(z; Y|x):=\go(Y|x)+W_1(z; Y)+W_2(z; Y)+\dots
\ee
The embedding is called {\it canonical} if
\be
W(0; Y|x)=\go(Y|x)\,.
\ee
The generating equations of \cite{Didenko:2023vna} read
\begin{align}
&\dr_x W+W*W=0\,,\label{xeq}\\
&\dr_z W+\{W,\Lambda\}_*+\dr_x\Lambda=0\,,\label{zeq}\\
&\dr_x C+\left(W(z'; y, \y)*C-C*W(z'; -y,
\y)\right)\Big|_{z'=-y}=0\,,\label{dCeq}
\end{align}
where $\dr_z=\dr z^{\al}\ff{\p}{\p z^{\al}}$, $C=C(y, \y)$ is $z$
independent just as it appears in \eqref{eqw}-\eqref{eqC}, and
\be\label{lmbd}
\Lambda(z; y, \y)=\dr
z^{\al}z_{\al}\int_{0}^{1}\dr\tau\tau\,e^{i\tau
z_{\gb}y^{\gb}}C(-\tau z, \y)
\ee
satisfies the condition
\be
\dr_z\Lambda=C(y, \y)*\gga\,,
\ee
where
\be
\gga=\ff12\, e^{iz_{\al}y^{\al}}\,\dr z^{\gb}\dr z_{\gb}\,,
\ee
while the star product $*$ extended to the $(z; Y)$ space has the
form
\be\label{limst}
(f*g)(z; Y)= \int f\left(z+u', y+u; \y \right)\star
g\left(z-v,y+v+v'; \y \right)
\exp({iu_{\al}v^{\al}+iu'_{\al}v'^{\al}})\,,
\ee
where $\star$ is a part of the star product \eqref{moyal} that
acts on $\y$ only,
\be\label{starvec}
(f\star g)(\y)= \int f\left(\y+{\mathbf{u}}\right)
g\left(\y+{\mathbf{v}} \right)
\exp({i{\mathbf{u}}_{\al}{\mathbf{v}}^{\al}})\,.
\ee
From the above integrations, it is easy to derive
\begin{align}
&y* =y+i\ff{\p}{\p y}-i\ff{\p}{\p z}\,,\qquad z* =z+i\ff{\p}{\p
y}\,,\\
&* y=y-i\ff{\p}{\p y}-i\ff{\p}{\p z}\,,\qquad * z=z+i\ff{\p}{\p
y}\,,\\
&\y* =\y+i\ff{\p}{\p \y}\,,\qquad * \y=\y-i\ff{\p}{\p \y}\,.
\end{align}
In particular, one observes that $z$'s commute
\be
[z_{\al}, z_{\gb}]_*=0\,.
\ee
Equation \eqref{limst} reduces to \eqref{moyal} for
$z$-independent functions. Equations \eqref{eqw}-\eqref{eqC}
result from \eqref{xeq}-\eqref{dCeq} order by order upon solving
for the $z$ dependence of $W$ using \eqref{zeq} and then
substituting the result into \eqref{xeq} for \eqref{eqw} and into
\eqref{dCeq} for \eqref{eqC}. However, the prescribed procedure
leads to the equations off shell. To set them on shell, one has to
choose representatives for $\go$ and $C$ and then strip the ideal
contribution from \eqref{eqw}-\eqref{eqC} off
\cite{Vasiliev:2003ev}. The ideal is generated with the help of
certain field-dependent $sp(2)$ generators found manifestly in
\cite{Didenko:2023vna}.

Let us point to an unusual property of the system
\eqref{xeq}-\eqref{dCeq}. The last equation \eqref{dCeq} of the
three is not independent. It follows from \eqref{zeq} via
consistency. This fact is not quite manifest, however. To check
it, one hits \eqref{zeq} with $\dr_z$ and uses the following
projective identity \cite{Didenko:2022qga}:
\begin{align}
&\dr_z(W(z; y, \y)*\Lambda)=(W(z'; y,
\y)*C)\Big|_{z'=-y}*\gga\,,\label{prid1}\\
&\dr_z(\Lambda*W(z; y, \y))=-(C*W(z'; -y,
\y))\Big|_{z'=-y}*\gga\,.\label{prid2}
\end{align}
Consistency of the equations \eqref{prid1}-\eqref{prid2} relies on
the specific star product \eqref{limst}, the precise form of
$\Lambda$ \eqref{lmbd}, and the functional class that evolves on
\eqref{xeq}-\eqref{zeq}, to which the field $W$ belongs. For more
details, we refer to \cite{Didenko:2022qga, Didenko:2023vna}. Let
us also note that the variable $z'$ within the argument of $W$
evades star multiplication and is set to $-y$, as prescribed
above.

\section{Solution}\label{secSol}

The natural vacuum of HS theory is $AdS_{d+1}$ space described by
a $z$-independent bilinear in a $Y$ flat connection $W_0$
satisfying \eqref{xeq}. It is convenient to choose it using the
Poincar\'{e} coordinates, in which the metric reads
\be
ds^2=\ff{1}{\rmz^2}(\dr \rmz^2+\eta^{ij}\dr \rmx_{i}\dr
\rmx_{j})\,,\label{Poincare}
\ee
where the radial coordinate $\rmz$ should not be confused with
$z_{\al}$, while $\rmx^j:=\vec\rmx$ are coordinates on the
$d$-dimensional Minkowski boundary, $i,j=0\dots d-1$ with metric
$\eta_{ij}$. Let us introduce the following notation for the
split-component $\y^a$:
\be
\y^a=\left\{ \begin{array}{ll} \y^j=\vec\y\,,\quad a=j<d\,, \\
i\bar y\,,\quad a=d
\end{array}\right.,
\ee
where the imaginary $i$ is introduced conventionally, while $\bar
y$ is not complex conjugate to any $\y$'s, but rather is an
independent component. The commutation relations
\be
[\y^{i}_{\al}, \y^{j}_{\gb}]_*=2\eta^{ij}\gep_{\al\gb}\,,\quad
[y_{\al}, y_{\gb}]_*=2i\gep_{\al\gb}\,,\quad [\bar y_{\al}, \bar
y_{\gb}]_*=-2i\gep_{\al\gb}
\ee
provide the following comprehensive set of $o(d,2)$ conformal
algebra generators
\be
M_{ij}=\ff12\y^{\al}_{i}\y_{j\al}\,,\quad
P_i=\ff12\y^{\al}_{i}(y-\bar y)_{\al}\,,\quad
K_i=\ff12\y^{\al}_{i}(y+\bar y)_{\al}\,,\quad D=-\ff12 y_{\al}\bar
y^{\al}\,.
\ee
The connection
\be\label{W0}
W_0=\ff{i}{\rmz}\left(\dr \rmx^{j}\,P_{j}-\dr \rmz \,D\right)
\ee
can be shown to satisfy \eqref{xeq}. We fix this vacuum in our
analysis by assuming that it receives no correction even in the
case of a nonzero field configuration of field $C$. In what
follows, we also need the associated star commutators derived from
\eqref{limst},
\begin{align}
&[\vec P, \bullet]_*=i\vec \y^\alpha\left(\frac{\partial}{\partial
\bar{y}^\alpha}+\frac{\partial}{\partial y^\alpha}\right)
-i\left(y-\bar y-i\frac{\partial}{\partial
z}\right)^{\al}\frac{\partial}{\partial
\vec\y^{\al}}\,,\label{Pcom}\\
&[D, \bullet]_*= -iy^\alpha \frac{\partial}{\partial
\bar{y}^\alpha}-i\bar{y}^\alpha \frac{\partial}{\partial
y^\alpha}+\epsilon^{\alpha \beta}\frac{\partial^2}{\partial
z^\alpha\, \partial \bar{y}^\beta}\,.\label{Dcom}
\end{align}

\paragraph{Weyl module and $T$ ansatz} The nontrivial part of the
following analysis is to solve for the Weyl module $C$ satisfying
\eqref{zeq} and \eqref{dCeq}. Let us specify conditions that we
impose to constrain our ansatz for $C$. First, as we have
mentioned, Eqs. \eqref{xeq}-\eqref{dCeq} result in no dynamics
unless on-shell representatives are picked and the trace ideal is
factored out. As shown in, e.g., \cite{Bekaert:2004qos}, a
convenient $C$-representative is {\it twisted}-traceless (due to
the twist \eqref{pi}), rather than the usual $AdS$ traceless
typical of the 1-form  $\go(Y|x)$. In our notation, the former
condition takes the form
\be\label{trC}
\Delta_{\al\gb}C:=\left(\eta^{ij}\ff{\p}{\p\y^{i\al}}\ff{\p}{\p\y^{j\gb}}-\ff{\p}{\p\bar
y^{\al}}\ff{\p}{\p\bar y^{\gb}}-y_{\al}y_{\gb}\right)C=0\,.
\ee
The operator $\Delta_{\al\gb}$ can be shown to commute with the
free equations arising from \eqref{dCeq} upon substituting
\eqref{W0} in place of $W$:
\be
\tilde{D} C=0\,,\qquad [\Delta_{\al\gb}, \tilde{D}]=0\,,
\ee
where
\be
\tilde{D}=\dr_x+\ff{\dr\vec\rmx}{\rmz}\cdot\left(i\vec\y^{\al}y_{\al}-
i\ff{\p}{\p\vec\y^{\al}}\ff{\p}{\p
y^{\al}}-\vec\y^{\al}\ff{\p}{\p\bar y^{\al}}-\bar
y^{\al}\ff{\p}{\p\vec\y^{\al}}\right)+i\ff{\dr\rmz}{\rmz}\left(
y_{\al}\bar y^{\al}+\gep^{\al\gb}\ff{\p}{\p y^{\al}}\ff{\p}{\p
\bar y^{\gb}}\right).
\ee
In addition, representatives singled out by \eqref{trC} respect
the action of the spin operator of the twisted-adjoint module
\be
\hat{s}=\ff12\left(\y^{a\al}\ff{\p}{\p\y^{a\al}}-y^{\al}\ff{\p}{\p
y^{\al}}\right)=\ff12(\vec\y\cdot\vec\p_{\al}-\bar
y^{\al}\bar\p_{\al}-y^{\al}\p_{\al})\,,\qquad [\hat s,
\Delta_{\al\gb}]=-\Delta_{\al\gb}\,.
\ee
Thus, Eq. \eqref{trC} is consistent with a natural choice of
having the scalar as the lowest component of the Weyl module
$C(y,\bar y; \vec\y|x)$, (see also \cite{Didenko:2012vh})
\be\label{scalar}
\phi(x)=C(0,0; \vec 0|x)\,.
\ee

Now, aiming at the exact solution, we would like $C$ to be nonzero
within the scalar sector only. This implies that the eigenvalue of
spin operator is zero,
\be\label{s0}
\hat s\, C=0\,.
\ee
Even though the scalar sources higher spins in interactions, in
general, this may not be the case for a highly symmetric scalar
profile. Thus, we assume $C$ to depend on the Poincar\'{e} radial
$\rmz$ only: in other words, we take it to be $\vec\rmx$
independent.

With the above preparations, we propose the following ansatz:
\be\label{ans}
C=\rmz^2 e^{iy_{\al}\bar y^{\al}}T(p,q; \rmz)\,,
\ee
where
\be\label{pq}
p=-\rmz^2\,\vec\y^{\al}\cdot\vec\y^{\gb}y_{\al}y_{\gb}\,,\qquad
q=2i\rmz^2\,y_{\al}\bar y^{\al}\,.
\ee
The normalization and $\rmz$-scaling are chosen conveniently.
Equation \eqref{s0} is trivially satisfied, justifying the scalar
structure of the module. The exponential factor in \eqref{ans}
although can be absorbed into $T$ remains conveniently isolated. A
similar exponential was already introduced in
\cite{Vasiliev:2012vf} as a kind of intertwining operator between
free bulk fields and boundary currents. At the lowest interaction
level, the same exponential was observed to commute with
nonlinearities of HS equations in their local form
\cite{Didenko:2017lsn, Didenko:2021vdb}. In addition, being a
star-product projector, it shares this characteristic feature with
the HS bulk-to-boundary propagators \cite{Didenko:2012vh},
\be\label{proj}
e^{iy\bar y}*e^{i y\bar y}=\ff14e^{i y\bar y}\,.
\ee
Given the universality of the above substitution in various
applications, we refer to \eqref{ans} as the {\it $T$ ansatz}.

\paragraph{Obtaining the solution} All is set to proceed with
solving \eqref{zeq} and \eqref{dCeq}. Plugging \eqref{W0} into
\eqref{zeq} and using that $W_0$ from \eqref{W0} is $z$
independent we arrive at two conditions for $\Lambda$ from the
$\dr\rmz$ and $\dr\vec\rmx$ 1-forms, correspondingly,
\begin{align}
&[D,
\Lambda_\al]_*=-i\rmz\ff{\p}{\p \rmz}\Lambda_\al\,,\\
&[P_i, \Lambda_{\al}]_*=0\,.\label{pL}
\end{align}
Equivalently, using \eqref{Pcom} and \eqref{Dcom}
\begin{align}
&\left(\bar y^{\gb}\ff{\p}{\p y^{\gb}}+ y^{\gb}\ff{\p}{\p \bar
y^{\gb}}+i\gep^{\gb\gga}\ff{\p}{\p z^\gb}\ff{\p}{\p\bar
y^{\gga}}\right)\Lambda_\al=\rmz\ff{\p}{\p \rmz}\Lambda_\al\,,\label{L1}\\
&\left(y-\bar y-i\ff{\p}{\p
z}\right)^{\gb}\ff{\p}{\p\vec\y^{\gb}}\Lambda_{\al}-
\vec\y^{\gb}\left(\ff{\p}{\p y}+\ff{\p}{\p\bar
y}\right)_{\gb}\Lambda_{\al}=0\,.\label{L2}
\end{align}
Substituting \eqref{ans} into \eqref{lmbd}
\be
\Lambda_{\al}=\rmz^2\,\int_{0}^{1}\dr\tau\tau\, z_{\al}e^{i\tau
z(y-\bar y)}\,T(\pp, \qq; \rmz)\,,\label{Tanz}
\ee
where we introduced
\be
\pp=-\tau^2\rmz^2\,\vec\y_{\al}\cdot\vec\y_{\gb}z^{\al}z^{\gb}\,,\qquad
\qq=-2i\tau\rmz^2\,z_{\al}\bar y^{\al}\,.
\ee
Plugging \eqref{Tanz} into \eqref{L1} and \eqref{L2} and after
quite a lengthy calculation (see the Appendix for a sketch of the
derivation) that repeatedly uses partial integration of the form
\begin{align}
z^{\al}\ff{\p}{\p z^{\al}}\int_{0}^{1}\dr\tau\rho(\tau)\,f(\tau
z)&=\int_{0}^{1}\dr\tau\rho(\tau)\,\tau\p_{\tau}f(\tau
z)=\nn\\
&=\rho(1)f(z)-\rho(\tau)\tau\,f(\tau z)\Big|_{\tau=0}-
\int_{0}^{1}\dr\tau\p_{\tau}(\tau\rho(\tau))f(\tau z)
\end{align}
we arrive at
\begin{align}
&\int_{0}^{1}\dr\tau\tau\,e^{i\tau z(y-\bar y)} \left(
T'_{\rmz}-4\rmz (T'_{\qq}+\pp T''_{\pp\qq}+\ff{\qq}{2}
T''_{\qq\qq}) \right)=0\,,\label{zcond}\\
&\int_{0}^{1}\dr\tau\tau^2\,e^{i\tau z(y-\bar
y)}(T'_{\qq}-3T'_{\pp}-2\pp T''_{\pp\pp}-\qq
T''_{\pp\qq})=0\,,\label{xcond}
\end{align}
where $T'_{s}=\ff{\p}{\p s}T$. At this stage, the on-shell
condition \eqref{trC} has not been imposed yet. It gives us
another constraint for our ansatz \eqref{ans}, namely,
\be\label{eq1}
d\cdot
T'_{p}+2p\,T''_{pp}-2T'_{q}-2\mathrm{z}^2 T''_{qq}=0\,,\qquad
d=\gd_{i}{}^{i}\,.
\ee
Let us now set
\begin{align}
&T'_{\rmz}-4\rmz (T'_{q}+p T''_{pq}+\ff{q}{2}
T''_{qq})=0\,,\label{eq2}\\
&T'_{q}-3T'_{p}-2p T''_{pp}-q T''_{pq}=0\label{eq3}
\end{align}
in order to satisfy \eqref{zcond} and \eqref{xcond}. Substituting
\eqref{ans} into \eqref{dCeq} gives no new conditions as it leads
again to \eqref{eq2} and \eqref{eq3}. This fact is not surprising
because, as was stressed (see also \cite{Didenko:2022qga}),
\eqref{dCeq} comes as a consistency condition of \eqref{zeq}.
Eventually, the solution we look for should satisfy the three
differential equations \eqref{eq1}-\eqref{eq3}. However, it should
be analytic in $p$ and $q$. The simplest solution of this system
is
\be\label{parts}
T=\nu=const\,,
\ee
which is not trivial; see \eqref{ans}. However, there is another
solution that can be found in terms of power series
\be\label{pwr}
T=\sum_{m, n}\ff{f_{m,n}(\rmz)}{m!n!}p^m\,q^n\,.
\ee
Substituting \eqref{pwr} into \eqref{eq1}-\eqref{eq3} and leaving
the technical details for the Appendix, let us present the final
result:
\be\label{solD}
T=\rmz^{d-4}\sum_{m,n}\left(\ff{1}{\rmz^2}\right)^{m+n}\ff{p^m
q^n}{m!n!\,\Gamma(2+2m+n)\,\Gamma\left(\ff {d-2}{2}-m-n\right)}\,.
\ee
The above power series can be summed up using the contour
representation of the gamma function
\be
\ff{1}{\Gamma(k)}=\oint\dr\rho\,\rho^{-k} e^\rho\,,\qquad k\in
\mathbb{N}
\ee
leading eventually to the following final form of $T$
\be\label{res}
T=\nu_1+\nu_2\,\oint
\dr\rho\,\ff{e^\rho}{\rho^2}\left(\rmz^2+\ff{p}{\rho^2}+\ff{q}{\rho}\right)^{\ff{d-4}{2}}\,,
\ee
where $\nu_1$ and $\nu_2$ are arbitrary constants, while $p$ and
$q$ are given by \eqref{pq}. The integration contour encircles the
origin to avoid branch cuts. Thus, the obtained vacuum of the
system \eqref{xeq}-\eqref{dCeq} is described by the connection
\eqref{W0} and Weyl module \eqref{ans} and \eqref{res}, which  we
present here in terms of the original variables for convenience:
\begin{align}
&W_0=\ff{i}{2\rmz}\left(\dr \rmx^{j}\,\y^{\al}_{j}(y-\bar
y)_{\al}+\dr \rmz \,y_{\al}\bar y^{\al}\right)\,,\label{Wvac}\\
&C=e^{iy_{\al}\bar
y^{\al}}\left(\nu_1\,\rmz^2+\nu_2\,\rmz^{d-2}\,\oint
\dr\rho\,\ff{e^\rho}{\rho^2}\left(1+\ff{x_1}{\rho^2}+\ff{x_2}{\rho}\right)^{\ff{d-4}{2}}
\right)\,,\label{Cvac}
\end{align}
where
\be
x_1=-\vec\y^{\al}\cdot\vec\y^{\gb}y_{\al}y_{\gb}\,,\qquad
x_2=2iy_{\al}\bar y^{\al}\,.
\ee
Assuming analyticity in $p$ and $q$, the solution \eqref{res} is
the only solution of the partial differential equations
\eqref{eq1}-\eqref{eq3}. This may sound surprising given no
boundary conditions were imposed, but in fact, as the analysis in
the Appendix shows, this system is somewhat fine-tuned to have
very few analytic solutions. For example, had the coefficient $p$
in \eqref{eq2} been different, say, $2p$, there would be no
analytic solutions at all, other than $T=const$.

\paragraph{Basic properties} Let us recapitulate some salient features of
the obtained background.
\begin{itemize}
\item The vacuum 1-form connection $W_0$ is given by the $AdS$
bilinears in the Poincar\'{e} coordinates \eqref{W0}. It remains
undeformed despite nontrivial scalar excitation in $C$. This in
turn implies that the scalar itself satisfies free equations
trivializing the nonlinear self-interaction. Its profile can be
extracted from \eqref{scalar},
\be\label{scl}
\phi(\vec\rmx, \rmz)=\nu_1\,\rmz^2+\nu_2\,\rmz^{d-2}\,.
\ee
It does not depend on the boundary coordinates $\vec\rmx$ and
offers an arbitrary mixture of its two branches\footnote{The case
of $d=4$ corresponding to $\Delta_1=\Delta_2$ is exceptional as
one loses the logarithmic scalar branch given by $\rmz^2\log\rmz$.
The missing branch cannot be captured by \eqref{res}. It would be
interesting to reconsider this case separately in particular to
see whether the free solution remains exact.} of conformal
dimensions $\Delta_1=2$ and $\Delta_2=d-2$. Given that the radial
coordinate $\rmz$ is dimensionless (in terms of the cosmological
constant), the two constants $\nu_1$ and $\nu_2$ carry the
standard dimension of a scalar in $d+1$. The solution is on shell
because it satisfies the chosen condition \eqref{trC} from factor
algebra. Since $W$ gains no $C$ corrections, and since the linear
$C$ is exact, our background  trivializes the higher-order
vertices \eqref{HO}.

\item The independence of Eq. \eqref{scl} from $\vec\rmx$ suggests
that the leftover global space-time symmetry of the vacuum is the
Poincar\'{e} algebra in $d$ dimensions spanned by the Lorentz
generators $M_{ij}$ and translations $P_i$. This is indeed the
case, as can be seen from the analysis of the conditions $\gd_\gep
W_0=\gd_\gep C=0$, which give
\begin{align}
&\dr_{x}\gep+[W_0, \gep]_*=0\,,\label{gep1}\\
&\dr_{z}\gep+[\Lambda, \gep]_*=0\,.\label{gep2}
\end{align}
Taking $\gep$ to be $z_{\al}$ independent, we have from
\eqref{gep2} that $[\Lambda, \gep]_*=0$. Taking into account
\eqref{pL} as well as the simple observation that Lorentz
generators commute with $\Lambda$, $[M_{ij}, \Lambda]_*=0$, the
space-time global symmetry parameter that satisfies \eqref{gep2}
reads
\be\label{glans}
\gep=\ff12\xi^{ij}(\vec\rmx, \rmz)M_{ij}+\xi^i(\vec\rmx,
\rmz)P_i\,,
\ee
where $\xi$ are some $x$-dependent parameters. Plugging
\eqref{glans} into \eqref{gep1}, it is easy to obtain that
$\xi_{ij}$ are arbitrary constants, while
\be
\xi_i=\ff{1}{\rmz}(\xi^0_{i}+\xi_{ji}\rmx^{j})\,,\qquad
\xi^0_i=const\,.
\ee
Thus, \eqref{glans} indeed parametrizes Poincar\'{e} algebra in
$d$ dimensions.

\item The structure of the $T$ module \eqref{res} is different in
odd and even dimensions. In the latter case, $T$ is always a
polynomial, e.g.,
\be
T_{d=8}=\nu_1+\nu_2\,\rmz^4\left(1+\ff{x_1^2}{120}+\ff{x_2^2}{6}+\ff{x_1}{3}+x_2+\ff{x_1x_2}{12}\right)\,,
\ee
while in the odd case it is not, as the integration in \eqref{res}
brings all powers of $p$ and $q$. Note also that the contour
representation \eqref{res} was already introduced in
\cite{Didenko:2012vh} in a different context. This integral can be
expressed in terms of the Gegenbauer polynomials
$C^{(\al)}_{n}(x)$ as follows:
\be
\oint \dr\rho
\ff{e^{\rho}}{\rho^2}\left(1-\ff{2xy}{\rho}+\ff{y^2}{\rho^2}\right)^{-\al}=\sum_n\ff{y^n}{(n+1)!}
C_{n}^{(\al)}(x)\,.
\ee
The Gegenbauer polynomials are known to arise as generating
functions of the conserved currents of the $O(N)$ model; see,
e.g., \cite{Anselmi:1999bb}. The presence of these polynomials in
the structure of the $T$ module may be a manifestation of the
Flato-Fronsdal theorem \cite{Flato:1978qz, Vasiliev:2004cm}.

\item One should be cautious about interpreting the fields
\eqref{Wvac} and \eqref{Cvac} as proper physical fields of the
on-shell system. Indeed, while they do enjoy the chosen
representative conditions and therefore are on shell, their
physical interpretation may not be straightforward, given that the
factorization procedure that brings the off-shell system
\eqref{xeq}-\eqref{dCeq} on shell is not yet detailed. It is
likely that the physical fields may acquire a form different from
\eqref{Wvac}-\eqref{Cvac}. That this might be the case is signaled
by the lack of corrections to the space-time background from a
scalar despite its nonzero stress tensor. Nevertheless, the
Poincar\'{e} symmetry of physical fields is guaranteed due to the
fact that the quotienting comes about in terms of the HS module
$C$ \cite{Didenko:2023vna}, which itself is Poincar\'{e} invariant
in our case.

\item As a final remark, let us stress the importance of the
Fock-type projector \eqref{proj} in the construction of the Weyl
module \eqref{ans}. It comes out in many HS applications. For
example, it offers a 'forgetful property' of the HS
bulk-to-boundary propagators, making the HS $N$-point correlators
calculable in \cite{Didenko:2012tv}. It also appears within the
structure of HS black holes; see, e.g., \cite{Didenko:2009td,
Iazeolla:2011cb, Didenko:2021vdb}. Most notably, in some cases it
makes consideration of the projector-based solutions within the
original Vasiliev framework of \cite{Vasiliev:1992av} problematic
due to the artificial star-product divergences it elicits at the
level of master fields. For example, in \cite{Aros:2017ror} a
class of various exact solutions that admit six isometries was
found. One of these (called type $DW_0$) is a four-dimensional
analog of our solution. It is also based on the Fock-type
projector, which, however, develops singularities to the lowest
interaction order. As a result, this particular solution of
\cite{Aros:2017ror} is supplemented with a specific regularization
prescription. Given that the vacuum obtained in this paper is free
from any divergences, we may expect that the divergences of
\cite{Aros:2017ror} are really spurious.
\end{itemize}

\section{Discussion}\label{secCon}
We constructed a very simple vacuum \eqref{Wvac}-\eqref{Cvac} of
the nonlinear bosonic HS theory in $d+1$ dimensions. All of its
fields vanish except for the scalar, which spreads along the
Poincar\'{e} radial direction $\rmz$ in $AdS$. Being highly
symmetric, it respects the Poincar\'{e} algebra that naturally
acts on $AdS$ slices at fixed $\rmz$ as the global space-time
symmetry. As a result, the obtained solution mildly breaks global
HS symmetry.

In obtaining this vacuum we chose a suitable $AdS_{d+1}$ flat
connection as a combination of translations and a dilatation from
the algebra $o(d,2)$ using the standard Poincar\'e coordinates. As
the HS equations \eqref{xeq}-\eqref{dCeq} are off shell, we were
forced to impose the extra condition \eqref{trC} that selects the
on-shell representative. To solve the system, the so-called $T$
ansatz \eqref{ans} based on the Fock projector was used. This
particular choice, first introduced at the free level in
\cite{Vasiliev:2012vf}, is motivated by the immunity of the Fock
projector to HS nonlinearities \cite{Didenko:2017lsn,
Didenko:2021vdb}. The $T$ module enjoys a system of partial
differential equations that admits an explicit solution in terms
of Gegenbauer polynomials. Quite remarkably, the solution we found
trivializes the interacting HS vertices \eqref{HO} in a given
frame, which makes it linearly exact. Thus, the scalar profile
features a superposition of its shadow $\Delta=2$ and current
$\Delta=d-2$ branches that come with arbitrary dimensionful
constants \eqref{scl}. In $d+1=4$ the analogous vacuum was earlier
found as a solution of Vasiliev's equations in
\cite{Aros:2017ror}, modulo regularized divergencies. The
formalism used in this paper features no divergences, thus
pointing at unphysical nature of the infinities
\cite{Aros:2017ror} clashed with.

An intriguing problem for the future is to elaborate on the
structure of the field spectrum about the proposed vacuum.
Naturally, we expect the corresponding theory to live in
$d$-dimensional flat space. It is conceivable that the spectra
differ for $\Delta=2$ and $\Delta=d-2$ vacua, as well as for the
mixture of the two. Given that the vacuum parameters $\nu_{1,2}$,
\eqref{scl} are dimensionful, one may expect the fluctuations on
the Minkowski space to acquire $\nu$-dependent masses, as an
option. Another feasible option is that the spectrum is massless,
while $\nu$ appears in interaction vertices. The latter case would
relate HS theory in $AdS$ to a hypothetical one in Minkowski space
in a way that infers no flat limit. In any case, the free-field
analysis does not promise to be immediately straightforward.
Indeed, while the constructed vacuum is on shell, the generating
system \eqref{xeq}-\eqref{dCeq} is not. This implies that one has
to factor out the ideal associated to the field traces to set free
fields on their mass shell. The form of this ideal is driven by
the field-dependent $sp(2)$ constructed in \cite{Didenko:2023vna}.
Unlike the case of the standard HS vacuum \eqref{stnd}, which
generates an ideal out of field traces at the free level, the new
vacuum makes the $sp(2)$ generators depend on the vacuum structure
of $T$ from \eqref{res}. This may offer a technical complication
in the process of on-shell factorization.

The vacuum obtained may also play an important role in the HS
$AdS/CFT$ correspondence \cite{Klebanov:2002ja, Sezgin:2003pt,
Leigh:2003gk, Giombi:2009wh}. So, in $d=3$, the case that sparked
a surge of interest due to its conjectural relevance to the
Wilson-Fisher model \cite{Klebanov:2002ja}, it is the spinorial
version of the solution \eqref{res} with $\nu_1=0$ that has
emerged as an intertwiner of fields and currents within the
Vasiliev equations at the free level \cite{Vasiliev:2012vf}. It
would be very interesting to extend this analysis to all orders.
In particular, the holomorphic (chiral) generating equations of
\cite{Didenko:2022qga} are accessible to all-order analysis. It is
also of interest to trace the $d=3$ HS symmetry breaking from the
boundary perspective in its massive state (see e.g.
\cite{Bardeen:2014paa}), where masses resulted from the breaking
of scale invariance. It is conceivable that the vacuum parameters
may be related to the mass of a scalar of the potential boundary
dual Poincar\'{e}-invariant quantum field theory.

\section*{Acknowledgments}
We are grateful to Simone Giombi, Ruslan Metsaev and especially
Per Sundell and Mikhail Vasiliev for useful discussions. We are
very thankful to the anonymous referee for very useful comments
and suggestions. VE would like to thank Yasha Neiman for his warm
hospitality at the OIST Quantum Gravity Unit while finishing this
work. VE acknowledges the financial support from the Foundation
for the Advancement of Theoretical Physics and Mathematics
``BASIS''.

\section*{Appendix: Derivation extras}
Here we show that \eqref{Tanz} solves Eq. \eqref{zeq} of the
generating system, which for the chosen $AdS$ connection
\eqref{W0} casts into \eqref{L1} and \eqref{L2}. To derive
\eqref{L1} and \eqref{L2} from \eqref{zeq} the following formulas
are very handy:
\begin{multline}
\mathbf{y}^\alpha_j (y_\alpha-\bar{y}_\alpha)\ast f(z,y,\bar{y},\vec\y)=\\
=\left(\mathbf{y}^\alpha_j+i\epsilon^{\alpha\beta}\frac{\partial}{\partial
\mathbf{y}^{\beta \, j}}\right) \left(y_\alpha
+i\frac{\partial}{\partial y^\alpha}-\frac{\partial}{\partial
z^\alpha}-\bar{y}_\alpha+i\frac{\partial}{\partial
\bar{y}^\alpha}\right)f(z,y,\bar{y},\vec\y)\, ,
\end{multline}
\begin{multline}
f(z,y,\bar{y},\vec\y)\ast \mathbf{y}^\alpha_j (y_\alpha-\bar{y}_\alpha)=\\
=\left(\mathbf{y}^\alpha_j-i\epsilon^{\alpha\beta}\frac{\partial}{\partial
\mathbf{y}^{\beta\, j}}\right)
\left(y_\alpha-i\frac{\partial}{\partial
z^\alpha}-i\frac{\partial}{\partial
y^\alpha}-\bar{y}_\alpha-i\frac{\partial}{\partial
\bar{y}^\alpha}\right)f(z,y,\bar{y},\vec\y)\,.
\end{multline}
Combining these gives us \eqref{Pcom}. One derives \eqref{Dcom}
analogously. Substituting \eqref{Tanz} into \eqref{L1} yields
\begin{multline}
\left[\mathbf{y}^\alpha_j(y_\alpha-\bar{y}_\alpha),\Lambda_\xi\right]_\ast=4\mathrm{z}^2
z_\xi z_\gamma \mathbf{y}^\gamma_j\times\\ \int_0^1 d\tau\, \tau^2
\, e^{i\tau z (y-\bar{y})}\Bigg\{\frac{\partial T}{\partial
\mathsf{q}}+\tau(1-\tau)(iz_\alpha(y^\alpha
-\bar{y}^\alpha))\frac{\partial T}{\partial \mathsf{p}}-4\tau
\frac{\partial T}{\partial \mathsf{p}}-2\tau
\mathsf{p}\frac{\partial^2 T}{\partial
\mathsf{p}^2}-\mathsf{q}\tau\frac{\partial^2 T}{\partial
\mathsf{p}\partial\mathsf{q}}\Bigg\}\,.
\end{multline}
To proceed, we notice that for $T=T(\mathsf{p}, \mathsf{q})$ we
have the identity
\begin{equation}\label{Euler}
\tau \frac{\partial T}{\partial \tau}=2\mathsf{p}\frac{\partial
T}{\partial \mathsf{p}}+\mathsf{q}\frac{\partial T}{\partial
\mathsf{q}}\,,
\end{equation}
which allows us to rewrite some terms as derivatives with respect
to $\tau$ and then integrate by parts using \eqref{parts}. This
way, we obtain
\begin{equation}
\left[\mathbf{y}^\alpha_j(y_\alpha-\bar{y}_\alpha),\Lambda_\xi\right]_\ast=4\mathrm{z}^2
z_\xi z_\gamma \mathbf{y}^\gamma_j\int_0^1 d\tau\, \tau^2 \,
e^{i\tau z_\alpha (y^\alpha-\bar{y}^\alpha)}\Bigg\{\frac{\partial
T}{\partial \mathsf{q}}-3 \frac{\partial T}{\partial
\mathsf{p}}-2\mathsf{p}\frac{\partial^2 T}{\partial
\mathsf{p}^2}-\mathsf{q}\frac{\partial^2 T}{\partial
\mathsf{p}\partial \mathsf{q}}\Bigg\}\,,
\end{equation}
which equals zero due to \eqref{eq3}.

In a similar way, the $\mathrm{dz}$ sector of \eqref{zeq} can be
solved. Plugging \eqref{Tanz} into \eqref{Dcom} and using
\eqref{Euler} gives
\begin{multline}\label{yyL}
\frac{i}{2\mathrm{z}}\left[y_\alpha \bar{y}^\alpha,\Lambda_\xi\right]_\ast=\mathrm{z} \, z_\xi\int_0^1 d\tau\,
\tau^2 \frac{\partial}{\partial \tau}\left(e^{i\tau z_\alpha
(y^\alpha-\bar{y}^\alpha)}\right)\left(1+2\mathrm{z}^2
\frac{\partial}{\partial \mathsf{q}}\right)T-\mathrm{z}\, z_\xi\int_0^1 d\tau\, \tau \,
e^{i\tau z_\alpha (y^\alpha -\bar{y}^\alpha)}\, \mathsf{q}\frac{\partial T}{\partial \mathsf{q}}-\\
-\mathrm{z}\, z_\xi \int_0^1 d\tau\, \frac{\partial}{\partial \tau}
\left(\tau^3\, e^{i\tau\, z_\alpha (y^\alpha- \bar{y}^\alpha)}
\left(1+2\mathrm{z}^2\frac{\partial}{\partial \mathsf{q}}\right)T\right)\,.
\end{multline}
Differentiation with respect to $\mathrm{z}$ then amounts to
\begin{equation}\label{dzL}
\frac{\partial \Lambda_\xi}{\partial \mathrm{z}}=z_\xi\int_0^1
d\tau \, \tau \, e^{i\tau \,
z_\alpha(y^\alpha-\bar{y}^\alpha)}\left(2\mathrm{z}+2\mathrm{z}\mathsf{p}\frac{\partial}{\partial
\mathsf{p}}+2 \mathrm{z}\mathsf{q}\frac{\partial }{\partial
\mathsf{q}}+\mathrm{z}^2 \frac{\partial}{\partial
\mathrm{z}}\right)T\,.
\end{equation}
Combining \eqref{yyL} and \eqref{dzL} one arrives at
\begin{multline}
\frac{\partial \Lambda_\xi}{\partial \mathrm{z}}+
\frac{i}{2\mathrm{z}}\left[y_\alpha \bar{y}^\alpha,\Lambda_\xi\right]_\ast=\\
=\mathrm{z}^2 \, z_\xi \int_0^1 d\tau\, \tau \, e^{i\tau \,
z_\alpha (y^\alpha-\bar{y}^\alpha)}\left(\frac{\partial T}{\partial \mathrm{z}}-
4\mathrm{z}\left(\frac{\partial T}{\partial \mathsf{q}}+
\mathsf{p}\frac{\partial^2 T}{\partial \mathsf{p}\,
\partial \mathsf{q}}+\frac{\mathsf{q}}{2}\frac{\partial^2 T}{\partial \mathsf{q}^2}\right)\right)\, ,
\end{multline}
which is again zero due to \eqref{eq2}.

\paragraph{Solving equations on $T$} Plugging the power series ansatz
\eqref{pwr} into \eqref{eq3} and \eqref{eq1}, we obtain the
following equations for coefficients, respectively:
\begin{equation}\label{eq5}
f_{m,n+1}=(3+2m+n)f_{m+1,n},
\end{equation}
\begin{equation}\label{eq6}
(d+2m)f_{m+1,n}-2 f_{m,n+1}-2\mathrm{z}^2 f_{m,n+2}=0.
\end{equation}
Using \eqref{eq5}, one can reduce \eqref{eq6} to
\begin{equation}
f_{m,n+1}=\frac{1}{\mathrm{z}^2}\dfrac{\frac{d}{2}-m-n-2}{2+2m+n}f_{m,n}\,\;\;\; \text{for}\;\;\; n\geq 1\,.
\end{equation}
This equation can be easily solved as
\begin{equation}
f_{m,n}(z)=\frac{\varphi(\mathrm{z})}{\mathrm{z}^{2(m+n)}}\frac{1}{\Gamma(2m+n+2)\Gamma(\frac{d}{2}-m-n-1)}\, ,
\end{equation}
where $\varphi(\mathrm{z})$ is a yet undefined function. Equation
\eqref{eq2} describes evolution with respect to $\mathrm{z}$ and
gives the following equation for $\varphi(\mathrm{z})$
\begin{equation}
\frac{\partial \varphi(\mathrm{z})}{\partial \mathrm{z}}=(d-4)\frac{\varphi(\mathrm{z})}{\mathrm{z}}\, .
\end{equation}
Thus, the general solution (up to an overall constant) of
equations on $f_{m,n}(\mathrm{z})$ reads
\begin{equation}
f_{m,n}(\mathrm{z})=
\frac{\mathrm{z}^{d-4-2m-2n}}{\Gamma(2m+n+2)\Gamma(\frac{d}{2}-m-n-1)}\,
.
\end{equation}
Plugging these coefficients back into the power series
\eqref{pwr}, one obtains \eqref{solD}.

\end{document}